# Observation of surface superconductivity in bulk polycrystalline MoS$_2$ induced by electric double-layer doping

Yoshihiro Shimazu[1,2,3]*, Tomonori Miyatake[1], Kento Ueno[1], and Masatomo Uehara[1]

[1]*Department of Physics, Yokohama National University, Hodogaya, Yokohama 240-8501, Japan*

[2]*Semiconductor and Quantum Integrated Electronics Research Center, Institute of Advanced Research, Yokohama National University, Hodogaya, Yokohama 240-8501, Japan*

[3]*Quantum Information Research Center, Institute of Advanced Sciences, Yokohama National University, Hodogaya, Yokohama 240-8501, Japan*

*E-mail: yshimazu@ynu.ac.jp

We report the observation of electric-field-induced superconductivity on the surface of bulk polycrystalline MoS$_2$ using electric double-layer doping. A gate voltage applied in an ionic liquid environment systematically increased carrier density, leading to an insulator-to-metal transition and a sharp resistance drop at low temperatures, indicating superconductivity. The onset temperature of superconductivity strongly depended on carrier density inferred from conductance, showing a significant increase and eventual saturation. Unlike prior studies limited to single-crystalline MoS$_2$, our results demonstrate that superconductivity can also be electrostatically induced in polycrystalline systems, broadening the scope for exploring gate-controlled superconductivity in a wider range of materials.

The electric-field control of carrier density has proven to be a powerful technique for building various electronic devices [1] and elucidating the physics of two-dimensional (2D) materials, such as graphene[2,3] and transition metal dichalcogenides (TMDCs).[4-8] Recently, electric double layers (EDLs) formed between material surfaces and ionic liquids (ILs) have garnered considerable attention because they allow very strong field effects,[9,10] enabling the emergence of novel electronic phases through EDL doping. A particularly intriguing subject is electric-field-induced superconductivity, whose realization has expanded to a growing range of materials.[11-18] The remarkable 2D nature of field-induced superconductivity presents a unique opportunity to explore the uncharted physics of 2D superconductivity and its potential applications.[19-21] Molybdenum disulfide ($MoS_2$), an archetypal layered semiconducting TMDC, has been extensively studied as an ideal platform of field-induced superconductivity. Single-crystalline $MoS_2$ and other TMDCs, in the form of monolayers, bilayers, and multilayer flakes, have been shown to exhibit superconductivity through electrostatic doping, mostly via EDL doping.[16,17,22-27] Recently, possible superconductivity was reported in CVD-grown polycrystalline monolayer $MoS_2$.[28] Tunneling spectroscopy of superconducting $MoS_2$ suggests that the field-induced superconducting state cannot be fully explained by the conventional BCS mechanism.[24,29] Instead, Ising superconductivity has been proposed.[27,29,30] Further investigation is needed to gain a deeper understanding of the mechanism of field-induced superconductivity. While superconductivity has been previously reported in single-crystalline $MoS_2$ with various thicknesses under electrostatic gating, the realization of superconductivity in bulk polycrystalline materials, including $MoS_2$, remains unexplored.

In this study, we report the first observation of electric-field-induced superconductivity in bulk polycrystalline $MoS_2$ using EDL doping. By applying a gate voltage $V_g$ in an IL environment, we modulated the carrier density and observed an onset of superconducting transition at low temperatures. Our results demonstrate that superconductivity in bulk $MoS_2$ can be induced purely by electrostatic means, without requiring chemical doping. Furthermore, we found that the superconducting transition temperature $T_c$ depends on the carrier density, as inferred from the conductance at 5 K. This behavior of $T_c$ is similar to that observed previously for $MoS_2$ [17,24] and $MoSe_2$.[22] This study was conducted as an initial step toward exploring electric-field-induced superconductivity in materials from which high-quality single crystals are difficult to obtain, such as Ni-based oxide $Ln_4Ni_3O_8$ (Ln = lanthanoid) — promising candidates for high-$T_c$ superconductivity [31-35]. By examining bulk polycrystalline $MoS_2$, we aimed to assess the feasibility of electrostatically inducing superconductivity in polycrystalline systems. Our findings indicate that superconducting transitions can be achieved in a range of bulk materials via electrostatic doping. This work opens new avenues for studying gate-controlled superconductivity on the surface of polycrystalline materials and may contribute to the development of novel superconducting electronic devices.

The disk-shaped polycrystalline pellets of $MoS_2$ (diameter: 10 mm, thickness: 1 mm) were prepared by pressing and sintering the powdered reagent of 2H-$MoS_2$ (99% up, Kojundo Chemical Laboratory Co., Ltd.) at 1000 °C for 20 h in a tubular furnace under flowing Ar gas. The samples were characterized by powder X-ray diffraction using Rigaku MiniFlexII with Cu-$K\alpha$ radiation, scanning electron microscopy, and energy dispersive spectroscopy. The XRD profiles shown in Fig. 1(a) indicate that the sample contains approximately 10 wt.% of $MoO_2$ as an impurity. The surface granularity observed by SEM is presented in Fig. 1(b).

The setup for EDL doping of the $MoS_2$ sample was devised as shown in Fig. 2(a). On one half of the disk-shaped sample (Fig. 2(b)), a 0.1-mm-thick Pt plate (Fig. 2(c)), serving as a gate electrode, was placed. Four holes (diameter: 1.8 mm) were used to insert spring-loaded pin contacts (PRECI-DIP 0906-1, plunger diameter: 1.07 mm). The pin contacts were held together using an acrylic plate and pushed onto the sample. The strength of the pushing force was adjusted using screws securing the acrylic plate. To prevent the pin contacts from touching the Pt plate, circular spacers made of Teflon, as shown in Fig. 2(c), were used. After injecting the IL, N, N-diethyl-N-(2-methoxyethyl) N-methylammonium bis-(trifluoromethanesulfonyl)-imid (DEME-TFSI), between the Pt plate and the sample, the sample was inserted into the cryostat and evacuated for over a day, reaching a pressure of approximately $10^{-4}$ Pa. Four-point resistance measurements were performed using a source-measure unit and a customized differential amplifier. At 240 K, above the freezing point of the IL, a gate voltage $V_g$ was applied to the Pt plate. The resistance gradually decreased with both increasing gate voltage and elapsed time. At a fixed $V_g$, the temperature dependence of resistance was measured down to 1.8 K. The four-point resistance was evaluated by sweeping the bias voltage from –10 mV to 10 mV or beyond. A linear relationship between the four-point voltage and the applied current (typically below 50 μA) was confirmed.

By holding the sample at 240 K with a fixed gate voltage $V_g$, the four-point resistance $R$ gradually decreased. Therefore, $V_g$ alone is not a sufficient parameter to characterize each dataset. A similar time-dependent resistance has been reported for exfoliated few-layer $MoS_2$ flakes,[24] attributed to the slow accumulation of carriers in the EDL formed by the IL at a given $V_g$.[24] Given the loosely packed polycrystalline nature of the present sample, a slow penetration of the IL into the surface is expected, which may also contribute to the observed time dependence of the resistance. Figures 3(a) and 3(b) show the temperature dependence of $R$ for various $V_g$ values and waiting times. To avoid possible chemical reactions at higher $V_g$, we limited the gate voltage to 6.6 V or below. The gate voltage conditions corresponding to Fig. 3(a) are summarized in Table 1. Datasets e, f, g, and h were sequentially acquired at $V_g$ = 6.6 V with varying waiting times. Subsequently, a large increase in resistance was achieved by applying a negative gate voltage of −6 V at 240 K. Datasets d, b, c, and a were then sequentially obtained, as detailed in Table 1. The reversibility and reproducibility of carrier control via gate voltage were confirmed.

Fig. 3(a) shows an insulator-to-metal transition with increasing carrier density. However, even at the highest carrier density—corresponding to the lowest curve (dataset h)—the resistance displayed a non-monotonic temperature dependence, showing a slight increase below 30 K. As shown in Fig. 3(b), the data acquired at $V_g$ = 0.1 and 5 V exhibited highly insulating behavior. Here, it is important to distinguish between the apparent carrier density and the local carrier density. Due to the porous and irregular surface morphology of the sample, the actual surface area is difficult to define precisely. The *apparent* sheet carrier density refers to the number of carriers per projected surface area of the sample, whereas the *local* sheet carrier density is defined on the actual surface area of individual grains. As a result, the apparent carrier density may substantially overestimate the local density. The observed insulator-to-metal transition clearly reflects an increase in local carrier density along with the apparent one.

Remarkably, a drop in resistance at low temperatures was observed for all datasets shown in Fig. 3(a), except for dataset a (i.e., the topmost curve). The small resistance drops for datasets b and c are highlighted in Fig. 4(a). Based on previous findings of superconductivity in EDL-doped $MoS_2$, this drop in resistance is attributed to the onset of the superconducting transition. Figure 4(a) shows the temperature-dependent resistance below 5.4 K, where the resistance is normalized to its value at 5 K and vertically offset for clarity. Notably, in datasets b, c, and d, the resistance first significantly increased monotonically before gradually dropping. A similar temperature dependence, i.e., a resistance drop following insulating behavior, was reported previously, and was attributed to weak localization [26] and significant disorder [25]. In our sample, strong disorder contributing to the insulating behavior above the superconducting transition may be attributed to grain boundaries. We found that the onset temperature of superconductivity decreased when the bias current exceeded 1 mA. However, this result is most likely due to a heating effect. Due to limitations in our equipment, a zero-resistance superconducting state, expected at lower temperatures, and the magnetic-field dependence of the superconducting transition were not observed.

The temperature dependence of resistance, shown in Fig. 4(a), clearly indicates that the onset temperature of superconductivity depends on the gating conditions. The onset temperature, $T_c$, was estimated by extrapolating straight lines, as illustrated in Fig. 4(a). Although the apparent or local carrier density could not be directly determined in this study, the conductance at 5 K, $G(5K)$, serves as a practical proxy. Figure 4(b) shows the dependence of $T_c$ on $G(5K)$, revealing a pronounced increase in $T_c$ followed by saturation. Similar saturation behavior has been reported for both thick (~20 nm) and thin (~3-layer) $MoS_2$ flakes [17,24]. A subsequent decrease in $T_c$ at higher carrier densities was also observed in Ref. [17]. In contrast, measurements on thinner $MoS_2$ flakes (<11 nm) on glass substrates showed that $T_c$ remains independent of carrier density [26]. The relationship between $T_c$ and carrier density remains an open question and is the subject of ongoing investigation [36-39] In this work, the saturated $T_c$ value of ~4.2 K is approximately half of that reported in earlier studies.[16,17,23,24,26,27] This reduction may suggest that the local carrier density did not reach sufficiently high levels. The observed saturation of $T_c$, along with the continuous decrease in resistance with increasing $G(5K)$, may reflect the saturation of local carrier density, despite a continued increase in apparent carrier density. Impurities inferred from XRD (Fig. 1(a)) and possible structural disorder in the sample may also contribute to the suppression of $T_c$ compared with single-crystalline $MoS_2$.

Another remarkable difference between the present work and previous studies[16,17,24] is the normal-state resistance just above $T_c$ (denoted as $R_N$) when $T_c$ saturation is observed. From Fig. 3(a), we find $R_N$ to be approximately 10 Ω at $T_c$ saturation, with the lowest $R_N$ observed being 3 Ω (dataset h). To compare these values with those reported in previous studies on single-crystalline samples,[16,17,24] the four-point resistance should be converted to sheet resistance. However, this conversion is not straightforward due to the granularity of the polycrystalline sample surface, as shown in Fig. 1(b). If we assume a flat surface, neglecting granularity, the sheet resistance is at most five times larger than the four-point resistance, given that the sample width is ~5 mm and the inner separation of the voltage contacts is ~1 mm. Thus, the estimated sheet resistance at $T_c$ saturation is less than 50 Ω—approximately two orders of magnitude smaller than those previously reported for thick (~20 nm) and thin (~3-layer) $MoS_2$ flakes.[17,24] In [16], a normal-state resistance of ~10 Ω was reported, but $T_c$ saturation occurred at $R_N$ > ~100 Ω, which is larger than the value found in this work. The low $R_N$ (apparent normal-state resistance) observed here could be attributed to the intrusion of the IL into the loosely packed polycrystalline sample. To estimate the critical carrier density for superconductivity, further investigation into the conductivity of the EDL-doped granular polycrystalline surface is necessary, as this aspect has not yet been studied. In principle, Hall-effect measurements could be performed on polycrystalline samples. However, such measurements yield the apparent sheet carrier density. Due to surface granularity, estimating the local sheet carrier density—necessary for meaningful comparison with single-crystalline samples—is not straightforward.

Another important consideration regarding the granular surface of polycrystalline samples is the increased surface area due to granularity.[40] If the gate electrode area in contact with the IL is significantly smaller than the sample area in contact with the IL, the doping efficiency decreases because the gate capacitance is dominated by the EDL capacitance of the gate electrode. Therefore, achieving a higher local carrier density requires a gate electrode with a much larger area than the sample surface area.[40] In our experimental setup, due to the sample's surface granularity, the gate electrode area is smaller than the total sample surface area exposed to the IL (see Fig. 2(a)). By modifying the experimental configuration—specifically, by using a significantly smaller sample—a higher local carrier density could be achieved. It is also worth noting that the granularity issues associated with bulk polycrystalline samples are not relevant to CVD-grown polycrystalline monolayer samples.[28]

In this study, we reported the observation of electric-field-induced superconductivity in bulk polycrystalline $MoS_2$ via EDL doping. Applying a gate voltage up to 6.6 V in an ionic liquid environment led to a systematic decrease in resistance, an insulator-to-metal transition, and a sudden resistance drop at low temperatures—strongly suggesting the onset of superconductivity in bulk $MoS_2$, which had not been observed before. These results indicate a successful increase in local carrier density at the surface. The superconducting onset temperature, $T_c$, strongly depended on the conductance at 5 K, $G(5\,K)$, which serves as a practical proxy for carrier density. $T_c$ increased significantly with $G(5\,K)$ and eventually saturated at ~4 K. This behavior is likely due to the saturation of local carrier density, even as the apparent carrier density continued to rise. Our findings provide new insights into the nature of electric-field-induced superconductivity in $MoS_2$, particularly in systems with possible structural disorder or granularity. They also demonstrate that electrostatic doping via an EDL can be effectively applied to materials from which high-quality single crystals are difficult to obtain. This approach broadens the range of candidate materials for tunable superconductivity—including both single-crystalline and polycrystalline systems—and offers promising opportunities for the development of novel electronic devices based on gate-controlled superconducting states.

**Acknowledgments**

This work was supported in part by JST Moonshot R&D program (JPMJMS2062).

**Table 1.** Gate voltage conditions (applied voltage and waiting time at 240 K) for the datasets shown in Fig. 3(a).

| Sequence | Dataset | Gate Voltage Condition |
|---|---|---|
| (1) | e | 6.6 V applied for 7.3 h before cooling |
| (2) | f | 6.6 V applied for 4.5 h before cooling |
| (3) | g | 6.6 V applied for 7.9 h before cooling |
| (4) | h | 6.6 V applied for 16.6 h before cooling |
| (5) | d | $V_g$ swept gradually from 6.6 V to -6 V over 2.2 h; then set to 0 V before cooling |
| (6) | i | -6 V applied for 0.9 h before cooling. Not shown in Fig. 3(a). |
| (7) | b | 4 V applied for 0.8 h; then set to 0 V before cooling |
| (8) | c | 5 V applied for 0.2 h; then set to 0 V before cooling |
| (9) | a | -6 V applied for 16.5 h before cooling |

**Figure captions**

**Fig. 1.** (a) Powder XRD profiles of MoS$_2$ reagent and sintered pellet. The intensity of the (002) peak is approximately five times the full scale of this graph. The sintered MoS$_2$ pellet exhibits the same 2H-MoS$_2$ phase as the reagent, although it contains a small amount of MoO$_2$ impurity. The impurity content is estimated to be around 10 wt.% based on the peak intensity ratio relative to MoS$_2$. (b) SEM micrograph of the polycrystalline sample, revealing a highly granular structure.

**Fig. 2.** (a) Side view of the experimental setup for EDL doping. Four spring-loaded pins are pressed onto the sample pellet. A 0.1-mm-thick Pt plate on the sample serves as a gate electrode. (b) Top view of the sample pellet. (c) The Pt plate serving as the gate electrode, with circular Teflon spacers shown.

**Fig. 3.** (a) Four-point resistance of the polycrystalline MoS$_2$ sample as a function of temperature. The carrier density was controlled by the applied gate voltage and the waiting time at 240 K. Datasets corresponding to different carrier densities are labeled with letters. The details of the gate voltage conditions are provided in Table 1. (b) Temperature dependence of the resistance at low carrier densities for $V_g$ = 0.1 V and 5 V, showing highly insulating behavior. The data for $V_g$ = 5 V were taken after a 1-hour wait at this gate voltage.

**Fig. 4.** (a) Temperature dependence of the resistance, normalized by its value at 5 K, for datasets b, c, d, f, g, and h shown in Fig. 3(a). These data are vertically offset for clarity. The onset temperature $T_c$ is defined as the intersection point of the extrapolated straight lines. All data indicate the onset of superconductivity. (b) The onset temperature of superconductivity as a function of the conductance (the inverse of the four-point resistance) at 5 K.

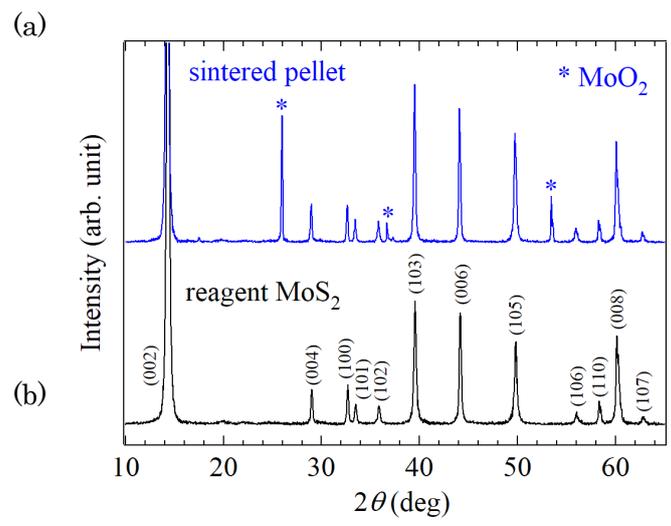

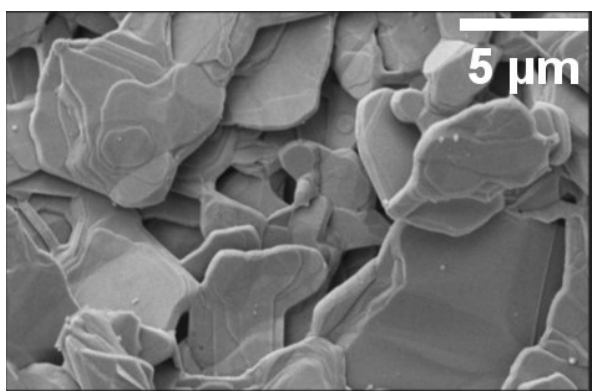

**Fig. 1.**

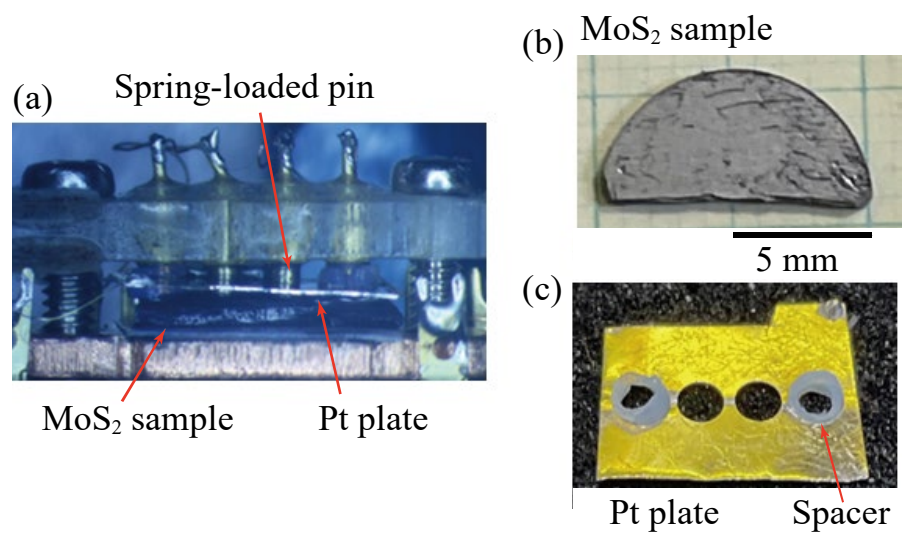

**Fig. 2.**

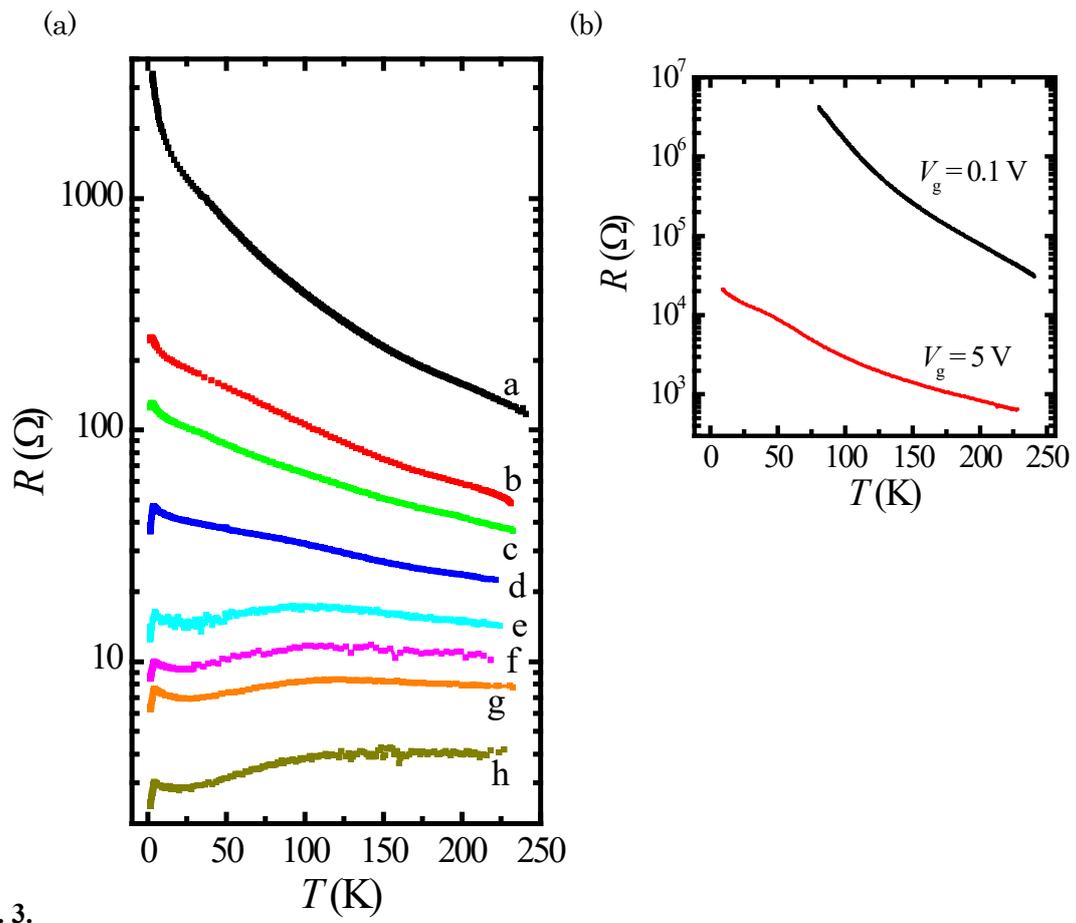

**Fig. 3.**

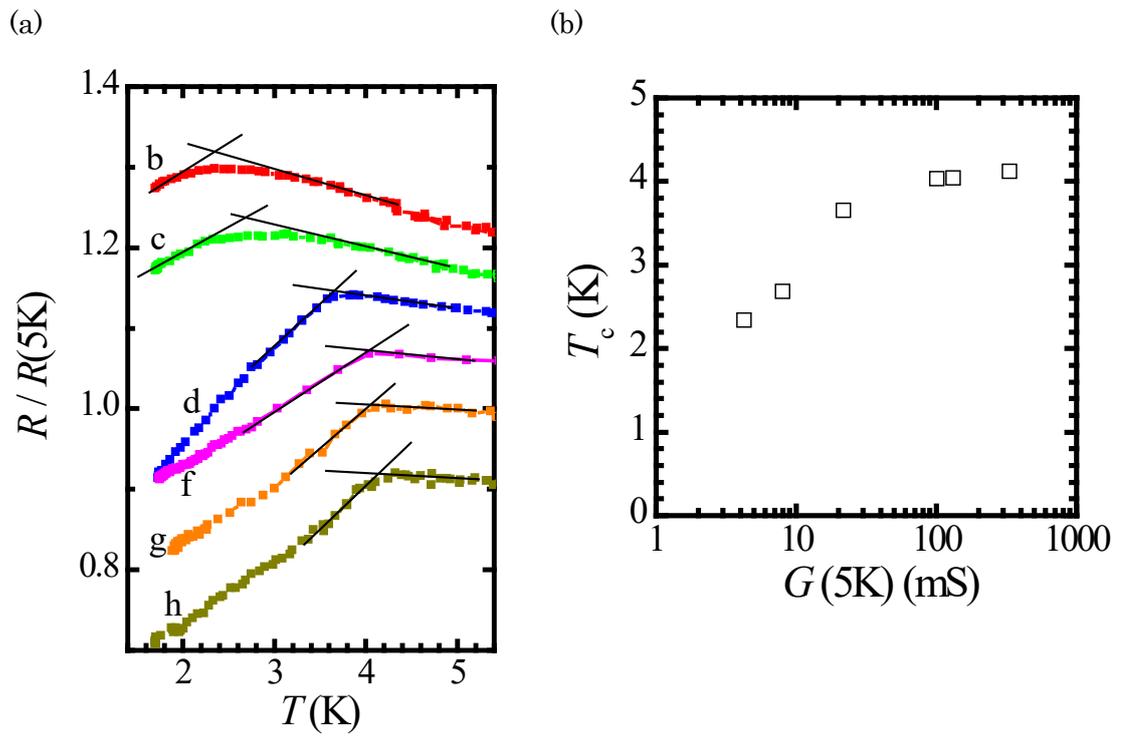

**Fig. 4.**